\documentstyle[12pt]{article}
\advance\hoffset by -7mm
\makeatletter
\@addtoreset{equation}{section}
\makeatother
\renewcommand{\theequation}{\thesection.\arabic{equation}}
\renewcommand{\title}[1]{\null\vspace{25mm}

\noindent{\Large{\bf #1}}\vspace{10mm}

\noindent {\large By }}
\newcommand{\authors}[1]{\noindent{\large #1}\vspace{20mm}

}
\newcommand{\address}[1]{\noindent #1\vspace{5mm}

}
\renewcommand{\abstract}[1]{\vspace{17mm}

\noindent{\small{\em Abstract.} #1}\vspace{2mm}

}
\newcommand{\journal}[4]{{\em #1~}{\bf #2}\,(19#3)\,#4;}

\newcommand{\hpa}{\journal {Helv. Phys. Acta}}

\newcommand{\ijmp}{\journal {Int. J. Mod. Phys.}}

\newcommand{\cmp}{\journal {Comm. Math. Phys.}}

\newcommand{\np}{\journal {Nucl. Phys.}}
\newcommand{\pl}{\journal {Phys. Lett.}}
\newcommand{\mpl}{\journal {Mod. Phys. Lett.}}
\newcommand{\prep}{\journal {Phys. Reports}}

\newcommand{\nc}{\journal {Nuovo Cim.}}

\newcommand{\lp}{\left(}\newcommand{\rp}{\right)}
\newcommand{\lc}{\left[}\newcommand{\rc}{\right]}
\newcommand{\lac}{\left\{}\newcommand{\rac}{\right\}}
\newcommand{\G}{\Gamma}
\newcommand{\D}{\Delta}
\renewcommand{\a}{\alpha}
\renewcommand{\b}{\beta}
\renewcommand{\d}{\delta}
\newcommand{\e}{\varepsilon}

\newcommand{\g}{\gamma}

\newcommand{\x}{\xi}
\renewcommand{\l}{\lambda} \renewcommand{\L}{\Lambda}
\newcommand{\m}{\mu}
\newcommand{\n}{\nu}

\renewcommand{\o}{\omega} \renewcommand{\O}{\Omega}

\newcommand{\r}{\rho}
\newcommand{\s}{\sigma} \renewcommand{\S}{\Sigma}
\newcommand{\th}{\theta}
\renewcommand{\t}{\tau}

\newcommand{\DD}{{\cal D}}

\newcommand{\LL}{{\cal L}}
\newcommand{\MM}{{\cal M}}
\newcommand{\NN}{{\cal N}}

\newcommand{\RR}{{\cal R}}
\newcommand{\SS}{{\cal S}}

\newcommand{\WW}{{\cal W}}

\newcommand{\complex}{{\kern .1em {\raise .47ex
\hbox {$\scriptscriptstyle |$}}
    \kern -.4em {\rm C}}}
\newcommand{\real}{{{\rm I} \kern -.19em {\rm R}}}
\newcommand{\rational}{{\kern .1em {\raise .47ex
\hbox{$\scripscriptstyle |$}}
    \kern -.35em {\rm Q}}}

\newcommand{\cb}{{\bar c}}

\newcommand{\pa}{\partial}

\newcommand{\fud}[2]{{\displaystyle{\frac{\delta #1}{\delta #2}}}}

\newcommand{\ie}{{{\em i.e.}\ }}

\newcommand{\sla}{\raise.15ex\hbox{$/$}\kern -.57em}

\newcommand{\twiddle}{\lower.9ex\rlap{$\kern -.1em\scriptstyle\sim$}}

\newcommand{\vf}{{\varphi}}

\newcommand{\equ}[1]{(\ref{#1})}

\newcommand{\eq}{\begin{equation}}
\newcommand{\eqn}[1]{\label{#1}\end{equation}}
\newcommand{\eea}{\end{eqnarray}}
\newcommand{\eqa}{\begin{eqnarray}}
\newcommand{\eqan}[1]{\label{#1}\end{eqnarray}}
\newcommand{\ba}{\begin{array}}
\newcommand{\ea}{\end{array}}
\newcommand{\eqac}{\begin{equation}\begin{array}{rcl}}
\newcommand{\eqacn}[1]{\end{array}\label{#1}\end{equation}}

\def\non{\nonumber\\}

\def\cb{\bar{c}}
\def\6{\partial}
\def\={\!\!\!&=&\!\!\!}
\def\+{\!\!\!&&\!\!\!+~}
\def\-{\!\!\!&&\!\!\!-~}

\def\ti{\tilde}
\def\ve{\varepsilon}

\renewcommand{\c}{\chi}
\newcommand{\tc}{\tilde{\chi}}
\newcommand{\fp}{\Phi\Pi}

\begin{document}

\setcounter{page}{0}
\thispagestyle{empty}
\hspace*{\fill} REF. TUW 95-05

\title{Local gravitational supersymmetry of Chern-Simons
theory in the vielbein formalism}
\authors{S. Emery\footnote{On leave from the D\'epartement de
Physique Th\'eorique, Universit\'e de Gen\`eve.
Supported by the ``Fonds Turrettini'' and the ``Fonds F. Wurth''.},
O. Moritsch\footnote{Work supported in part by the
         ``Fonds zur F\"orderung der Wissenschaftlichen Forschung''
         under Contract Grant Number P9116-PHY.},
M. Schweda,
T. Sommer\footnote{Work supported in part by the
         ``Fonds zur F\"orderung der Wissenschaftlichen Forschung''
         under Contract Grant Number P10268-PHY.}
and H. Zerrouki$^3$}
\address{Institut f\"ur Theoretische Physik,
         Technische Universit\"at Wien\\
         Wiedner Hauptstra\ss e 8-10, A-1040 Wien (Austria)}
\begin{flushleft}
March 1995
\end{flushleft}
\abstract{We discuss the Chern-Simons theory in three-dimensional
curved space-time in the vielbein formalism. Due to the additional
presence of the local Lorentz symmetry, beside the diffeomorphisms,
we will include a local gravitational supersymmetry
(superdiffeomorphisms and super-Lorentz transformations), which
allows us to show the perturbative finiteness at all orders. }

\centerline{\Large {}}


\newpage

\section{Introduction}

Topological gauge theories\footnote{See~\cite{blau1}
for a general review.},
at least in a Landau type gauge\footnote{Some non-covariant
gauges, like the axial gauge, are also
possible~\cite{andi1,andi2,stephane}.},
possess an interesting kind of symmetry,
namely the linear vector supersymmetry~\cite{oli}.
This feature has been extensively discussed by several groups
for a whole class of topological field theories, including the BF
models~\cite{ader,wallet,maggiore,silvio,hassan}, the bosonic and
the fermionic string, respectively in the Beltrami and
super-Beltrami parametrization~\cite{stora,becchi2,werneck,boresch},
four-dimensional topological Yang-Mills theory~\cite{oliveira,oti},
etc.

Another interesting topological example is the Chern-Simons
theory~\cite{blau1,schwarz,witten,king,axelrod} in three-dimensional
flat space-time formulated in the Landau gauge. Indeed, it is
invariant under a set of supersymmetry transformations whose
generators form
a Lorentz three-vector~\cite{delduc,birmingham1,olivier,olivier2}.
There, the generators of BRS transformations, supersymmetry and
translations obey a graded algebra of the Wess-Zumino type, which
closes on the translations. With the help of this supersymmetric
structure one is able to show the perturbative finiteness of such a
theory~\cite{olivier2,sorella}.

On an arbitrary curved space-time three-manifold, a local version of
this supersymmetry for the Chern-Simons theory in the Landau
gauge was derived in~\cite{olivier} and it has also been shown that,
in the perturbative approach, this local supersymmetry is
anomaly-free and that the theory is UV finite.
In~\cite{olivier}, the Einstein
formalism was used to describe the explicit metric dependence of the
gauge-fixing term\footnote{Remark that although the invariant
Chern-Simons action is topological, the gauge-fixing is not
metric independent.}. Starting from the requirement of invariance
under diffeomorphisms, a local supersymmetry for diffeomorphisms has
been constructed, which together with the BRS transformations form a
closed graded algebra.

In the present paper we extend the discussion of~\cite{olivier}
in such a way that we are
using the vielbein formalism to describe the local version of the
Chern-Simons theory on a three-manifold.
This allows us to incorporate also torsion.
Then, instead of only imposing diffeomorphism
invariance, we demand invariance of the action under gravity
transformations, which include beside the diffeomorphisms also local
Lorentz rotations. This generalization of the approach leads to a
local gravitational supersymmetry, which contains both
superdiffeomorphisms and super-Lorentz transformations.

This work is organized as follows.
In section 2 we briefly introduce the vielbein formalism and describe
the geometry with the Cartan structure equations. An important fact
is the independence of the model with respect to
the affine spin connection and the torsion. We will also formulate
the local theory by demanding the invariance under local gravity
transformations.
It has been shown that the ``physical'' content of the theory is
metric independent~\cite{witten,chen,asorey,coste,blau2,abud},
because the vielbein (or the metric) plays the role of a gauge
parameter.
In order to control the vielbein dependence, which follows from the
Landau gauge fixing, we extend the
BRS transformations~\cite{olivier,sibold}
and introduce the vielbein in a BRS doublet, which guarantees its
non-physical meaning.

The discussion of the symmetry content of that local theory
will follow in section 3. As it will be shown, beside the
superdiffeomorphisms, a further symmetry exists, which will be called
super-Lorentz transformations, and both can be combined to give a
local gravitational supersymmetry,
which breaks the gauge-fixed action at the functional level.
Indeed, the corresponding Ward identity contains a ``hard'' breaking
term, \ie a non-linear term in the quantum fields.
Fortunately, we are able to control this breaking by
means of a standard procedure~\cite{symanzik} using the fact, that the
breaking is a BRS-exact term. Thus, we enlarge the BRS transformations
by introducing a set of external fields grouped in a BRS doublet.
This allows us to present the local
gravitational supersymmetry in terms of unbroken Ward identities.
The Ward operator of this supersymmetry, together with the Ward
operator of gravity and the linearized Slavnov operator, form then
a closed linear graded algebra.

Section 4 is devoted to the study of the stability and the finiteness
of the local Chern-Simons theory.
The whole set of constraints (gauge condition,
ghost equation, antighost equation, Slavnov identity,
Ward identities for gravity and gravitational supersymmetry), which
are fulfilled by the classical action, completely fixes the total
action, \ie it forbids any deformations of it.
Furthermore, it will be shown that all the symmetries are free of
anomalies.
The stability of the classical theory, together
with the possibility of extending the classical constraints
to all orders of perturbation theory, ensures UV finiteness of
the quantum theory~\cite{olivier}.

In the appendix one finds an analysis with respect to
the trivial counterterms and their parameter independence.


\section{Local Chern-Simons theory in curved space-time at the
classical level}

The classical invariant Chern-Simons action $\S_{CS}$ can be written
in the space of forms according to\footnote{The wedge product
$\wedge$ has to be understood in the space of forms.}
\eq
\S_{CS}=-\frac{1}{2}\int_{\MM}\lp A^{A}dA^{A}
+\frac{\l}{3}f^{ABC}A^{A}A^{B}A^{C}\rp \ ,
\eqn{SIGMA_CS_1}
\centerline{}
where the gauge field $A^{A}=A^{A}_{\m}dx^{\m}$ has form degree one
and $d=dx^\m\6_\m$ denotes the exterior derivative.
The coupling constant $\l$ is related to the parameter $k$ by
$\l^2=2\pi/k$ (see~\cite{witten,king,olivier}).
We also assume that the gauge group is compact and that all fields
belong to its adjoint representation with $f^{ABC}$
as structure constants\footnote{Gauge group indices
are denoted by capital Latin letters $(A,B,C,...)$ and refer
to the adjoint representation, $[G^{A},G^{B}]=f^{ABC}G^{C}$,
$Tr(G^{A}G^{B})=\d^{AB}$.}.

As usual, one has to fix the gauge.
We choose a Landau type gauge fixing procedure. For the curved
space case we consider here, this gauge fixing term will be
implemented by means of the well-known Cartan approach.
This formalism allows us to
describe a general Riemannian manifold $\MM$ with the help of the
vielbein $e^a_\m (x)$ and the affine spin connection
$\o^{ab}_{~~\m}(x)$.
The endowed metric $g_{\m\n}(x)$ on $\MM$ in local coordinates
can be decomposed as
\eq
g_{\m\n}=e^a_\m e^b_\n\eta_{ab} \ ,
\eqn{METRIC}
where $\eta_{ab}$ denotes the Euclidean metric in the tangent
space\footnote{The tangent space indices $(a,b,c,...)$ are referred
to the group $SO(3)$.}. In order to describe the inverse metric
$g^{\m\n}(x)$ one needs the inverse vielbein $E_a^\m (x)$ which is
related to the vielbein by
\eq
e^{a}_{\mu}E^{\mu}_{b} = \d^{a}_{b}~~~,~~~
e^{a}_{\mu}E^{\nu}_{a} = \d^{\nu}_{\mu} \ .
\eqn{VIEL_ORTHO}
Therefore, $g^{\m\n}(x)$ takes the form
\eq
g^{\m\n}=E_a^\m E_b^\n\eta^{ab} \ ,
\eqn{INV_METRIC}
In order to describe symmetry transformations, one introduces also
the affine spin connection $\o^{ab}_{~~\m}$, which
is antisymmetric in the indices $(ab)$. Both, the vielbein and the
affine spin connection can be written in terms of forms according to
\eq
e^a=e^a_\m dx^\m~~~,~~~\o^{ab}=\o^{ab}_{~~\m}dx^\m
=\o^{ab}_{~~m}e^m \ .
\eqn{BASIC_FORM_1}
Finally, we are able to write down the Cartan structure equations
which are defined as
\eqa
T^a\=de^a+\o^{a}_{~b}e^b=\frac{1}{2}T^a_{\m\n}dx^\m dx^\n
=\frac{1}{2}T^a_{mn}e^m e^n \ , \\
R^{a}_{~b}\=d\o^{a}_{~b}+\o^{a}_{~c}\o^{c}_{~b}
=\frac{1}{2}R^a_{~b\m\n}dx^\m dx^\n=\frac{1}{2}R^a_{~bmn}e^m e^n \ ,
\eqan{CARTAN_STRUCT}
with the torsion 2-form $T^a$ and the curvature 2-form $R^{a}_{~b}$.

The Landau type gauge-fixing takes the following form
\eq
\S_{gf}=-s\int d^{3}\!x~\lc eE_a^{\m}E^{\n a}
(\6_{\n}\bar{c}^{A})A^{A}_{\m}\rc \ ,
\eqn{GAUGE_FIX}
where $e=det(e^a_\m)$ and $E_a^\mu=E_a^\mu(e^b_\n)$.
With $s$ as the nilpotent BRS operator,
the BRS transformations are given by
\eqa
\label{BRS_A}
sA^{A}\= Dc^{A}=dc^{A}+\l f^{ABC}A^{B}c^{C} \ , \\
\label{BRS_C}
sc^{A}\= \frac{\l}{2}f^{ABC}c^{B}c^{C} \ , \\
s\bar{c}^{A}\=B^A \ , \\[2mm]
sB^A\=0 \ .
\eqan{BRS_1}
One observes that the gauge-fixing term depends explicitely on
the vielbein field and therefore represents an inavoidable
non-topological contribution.
An important fact is, that the vielbein $e^a$
plays the role of a gauge parameter.
In order to guarantee its non-physical meaning, we let it transform
as a BRS doublet~\cite{olivier,sibold}:
\eqa
se^a\=\hat{e}^a \ , \\
s\hat{e}^a\=0 \ .
\eqan{BRS_DOUBLETS}

At this point let us remark that the vielbein, the
affine spin connection and the torsion, allow for a
complete tangent space formulation of the Chern-Simons action which
reads
\eqa
\S_{CS}=-\frac{1}{2}\int d^{3}\!x~ \ve^{mnk}\lc A^A_m\6_n A^A_k
+A^A_m A^A_l \lp\frac{1}{2}T^l_{nk}-\o^l_{~kn}\rp
+\frac{\l}{3}f^{ABC}A^A_m A^B_n A^C_k \rc ,&&
\eqan{ACTION_TANGENT}
where $\ve^{mnk}$ denotes the totally antisymmetric tensor with
indices in the tangent space\footnote{Through our definition
$\ve^{mnk}=e^m_\m e^n_\n e^k_\r \ve^{\m\n\r}$, with $\ve^{\m\n\r}$
as the totally antisymmetric contravariant tensor density with
weight 1, $\ve^{mnk}$ is a scalar density with weight 1 under
diffeomorphisms.}.
On the other hand, by introducing the covariant derivative
$\nabla_\m$ with respect to the affine connection $\G^\r_{\m\n}$,
which acts on an arbitrary covariant vector field in the usual
way~\cite{sexl}
\eq
\nabla_\m X_\n = e^a_\n \DD_\m (E_a^\r X_\r)
= \6_\m X_\n -\G^\r_{\m\n} X_\r \ ,
\eqn{NABLA}
the invariant Chern-Simons action \equ{SIGMA_CS_1} can be
rewritten as follows:
\eq
\S_{CS}=-\frac{1}{2}\int d^{3}\!x~\ve^{\m\n\r}
\lp A^A_\m \nabla_\n A^A_\r
+ \frac{1}{2}A^A_\m A^A_\l E^\l_l T^l_{\n\r}
+\frac{\l}{3}f^{ABC}A^A_\m A^B_\n A^C_\r \rp \ .
\eqn{ACTION_COVDER}
Due to the presence of the totally antisymmetric contravariant tensor
density $\ve^{\m\n\r}$ with weight 1,
the symmetric part of the affine connection vanishes and
the antisymmetric part of it will be canceled by the torsion term,
\ie
\equ{ACTION_COVDER} reduces to
\eq
\S_{CS}=-\frac{1}{2}\int d^{3}\!x~\ve^{\m\n\r}
\lp A^A_\m \6_\n A^A_\r
+ \frac{\l}{3}f^{ABC}A^A_\m A^B_\n A^C_\r \rp \ .
\eqn{ACTION_WORLD}
All three formulations \equ{ACTION_TANGENT},
\equ{ACTION_COVDER}
and \equ{ACTION_WORLD} are completely equivalent to the original
Chern-Simons action \equ{SIGMA_CS_1}, which implies that the model is
independent of the affine spin-connection $\o$ and the torsion $T$.

In order to translate the BRS invariance of the gauge-fixed action
into a Slavnov identity, one has to
couple the non-linear parts of the BRS transformations \equ{BRS_A}
and \equ{BRS_C} to external sources $\g^{\m A}$ and $\t^A$.
In terms of forms, these two external sources correspond to the dual
2-form $\tilde{\g}^A$ and the dual 3-form $\tilde{\t}^A$
\eqa
\ti{\g}^{A}\=\frac{1}{2}\ve_{\m\n\r}\g^{\r A}dx^{\m}dx^{\n} \ , \\
\ti{\t}^{A}\=\frac{1}{6}\ve_{\m\n\r}\t^{A}dx^{\m}dx^{\n}dx^{\r} \ ,
\eqan{DUALFORMS_1}
where $\ve_{\m\n\r}$ is the totally antisymmetric covariant tensor
density with weight -1.
This implies that $\g^{\m A}$ is a contravariant vector density
and $\t^{A}$ is a scalar density, both with weights 1.
\noindent
The contribution from the external sources to the complete action
hence writes in terms of forms
\eq
\S_{ext}=\int_{\MM}\lp \ti{\g}^{A}Dc^{A}
+\frac{\l}{2}f^{ABC}\ti{\t}^{A}c^{B}c^{C} \rp \ ,
\eqn{SIGMA_EXT_1}
or in components
\eq
\S_{ext}=\int d^{3}\!x~\lp -\g^{\mu A}D_\mu c^{A}
+\frac{\l}{2}f^{ABC}\t^{A}c^{B}c^{C} \rp \ ,
\eqn{SIGMA_EXT_2}
with the BRS transformations
\eq
s\ti{\g}^{A}=s\ti{\t}^{A}=0 \ .
\eqn{BRS_SOURCES}
\noindent
So far, the total Chern-Simons action in the Landau gauge with
external sources
\eq
\S=\S_{CS}+\S_{gf}+\S_{ext}
\eqn{ACTION_1}
is invariant under the BRS transformations\footnote{The total
action \equ{ACTION_1} is by construction invariant
under diffeomorphisms and local Lorentz transformations.}.

At the functional level, this
invariance is implemented by means of the Slavnov identity
\eq
\SS(\S)=\int d^{3}\!x~\lp \frac{\d\S}{\d{\g}^{\mu A}}
\frac{\d\S}{\d A^A_\m}+\frac{\d\S}{\d{\t}^A}\frac{\d\S}{\d c^A}
+B^A\frac{\d\S}{\d\bar{c}^A}
+\hat{e}^a_\m\frac{\d\S}{\d e^a_\m} \rp = 0 \ ,
\eqn{SLAVNOV_1}
and the corresponding linearized Slavnov operator
$\SS_{\S}$ writes
\eqa
\SS_{\S}=\int d^{3}\!x~\lp \frac{\d\S}{\d{\g}^{\m A}}
\frac{\d}{\d A^A_\m}+\frac{\d\S}{\d A^A_\m}\frac{\d}{\d{\g}^{\m A}}
+\frac{\d\S}{\d{\t}^A}\frac{\d}{\d c^A}
+\frac{\d\S}{\d c^A}\frac{\d}{\d{\t}^A}
+ B^A\frac{\d}{\d\bar{c}^A}
+\hat{e}^a_\m\frac{\d}{\d e^a_\m} \rp .~&&
\eqan{SLAVNOV_OPER_1}
Moreover, the classical action $\S$ \equ{ACTION_1} fulfills the
gauge condition
\eq
\frac{\d\S}{\d B^A}=\6_\m(eE^\m_a E^{\n a} A^A_\n) \ ,
\eqn{GAUGE_CONDITION}
and the ghost equation of motion
\eq
\lp \frac{\d}{\d\bar{c}^A}+\6_\m(eE^\m_a E^{\n a}
\frac{\d}{\d\g^{\n A}}) \rp \S
= -\6_\m \lp s(eE^\m_a E^{\n a}) A^A_\n \rp \ ,
\eqn{GHOST_EQU}
which one easily finds by (anti-)commuting the gauge condition
with the Slavnov identity~\cite{olivier2}.
The homogenous ghost equation implies that the
external field $\g^{\m A}$ and the antighost $\bar{c}^A$ can
only appear through the combination
$\chi^{\m A}=\g^{\m A}+eE_a^{\m}E^{\n a}\6_{\n}\bar{c}^{A}$,
which when written as a dual 2-form is given by
\eq
\tilde{\chi}^A=\tilde{\g}^A+\frac{1}{2}\ve_{\m\n\r}
eE_a^{\r}E^{\s a}(\6_{\s}\bar{c}^{A})dx^\m dx^\n \ .
\eqn{COMBINATION}
In addition, the action \equ{ACTION_1} fulfills a further global
constraint, namely the antighost equation~\cite{blasi}
\eq
\int d^{3}\!x~\lp \frac{\d\S}{\d c^A} +\l f^{ABC}\cb^B
\frac{\d\S}{\d B^C}\rp = \int_{\MM}\l f^{ABC}
\lp\tilde{\g}^B A^C + \tilde{\tau}^B c^C\rp \ .
\eqn{ANTIGHOST}
Furthermore,~\equ{ACTION_1} possesses other symmetries of the
gravitational type which will be discussed in the next section.


\section{Gravitational type symmetries of the classical action}

In the last paragraph we have stated the complete action
\equ{ACTION_1} for the Chern-Simons model in the vielbein formalism.
We showed that such a formulation allows us to consider the general
case for a Riemannian Manifold $\MM$.
We also exhibit its BRS invariance.
Moreover, this model contains a larger class of invariances.
Indeed, in the Einstein formalism, $\S$ is invariant under the
diffeomorphisms and the superdiffeomorphisms~\cite{olivier}.
In the Cartan framework, one has in addition also
the local Lorentz symmetry which describes the invariance of the theory
under local rotations. At this point, it is legitim to ask if there
exists another symmetry beside the superdiffeomorphisms. As one will
see, such a symmetry exists and will be called local super-Lorentz
transformations throughout this paper.

In what follows, diffeomorphisms and local Lorentz transformations will
be combined in the so-called local gravity transformations.
Analogously, the superdiffeomorphisms and the super-Lorentz symmetry
will generate a set of local gravitational supersymmetry
transformations\footnote{This local gravitational supersymmetry has
nothing to do with the ordinary Wess-Zumino type
supersymmetry~\cite{wz}.}

The gravity transformations of the elementary fields
$\vf=A,c,\cb,B,\g,\t,e,\hat{e}$
are given by:
\eq
\d^{gr}_{(\e,\O)}\vf = \LL_{\e}\vf + \d_{\O}\vf \ ,
\eqn{GRAV_TRANS}
where $\e^\m$ and $\O^{ab}$ are the infinitesimal parameters
of diffeomorphisms and local Lorentz transformations respectively,
both with ghost number +1.
The symbol $\LL_{\e}$ denotes the usual Lie
derivative in the direction of $\e^\m$ and $\d_{\O}$ describes the
Lorentz rotation operator.
At the functional level, the invariance of the classical
action~\equ{ACTION_1} under gravity transformations is expressed by
a Ward identity
\eq
\WW^{gr}_{(\e,\O)}\S=0 \ ,
\eqn{WARD_GRAV_ID}
where $\WW^{gr}_{(\e,\O)}$ denotes the Ward operator
\eq
\WW^{gr}_{(\e,\O)}=\int d^{3}\!x~\sum_{\vf}
\lp\d^{gr}_{(\e,\O)}\vf\rp
\frac{\d}{\d\vf} \ .
\eqn{WARD_GRAV}

Concerning the aforementioned local gravitational supersymmetry,
let us consider the following infinitesimal transformations:
\eq
\begin{array}{ll}
{}&\\
\d^{S}_{(\x,\th)}c^A=-\x^\m A^A_\m \ ,~~~    &
\d^{S}_{(\x,\th)}A^A_\m=\ve_{\m\n\r}\x^\n\chi^{\r A} \ , \\[4mm]
\d^{S}_{(\x,\th)}\chi^{\m A}=\d^{S}_{(\x,\th)}\g^{\m A}
=-\x^\m \t^A \ ,~~~    &
\d^{S}_{(\x,\th)}\t^A=0 \ ,  \\[4mm]
\d^{S}_{(\x,\th)}B^A=\d^{gr}_{(\x,\th)}\cb^A \ ,~~~    &
\d^{S}_{(\x,\th)}\cb^A=0 \ , \\[4mm]
\d^{S}_{(\x,\th)}\hat{e}^a_\m=\d^{gr}_{(\x,\th)}e^a_\m \ ,~~~    &
\d^{S}_{(\x,\th)}e^a_\m=0 \ , \\[4mm]
\end{array}
\eqn{SUPERGRAV}
where $\x^\m$ and $\th^{ab}$ are the infinitesimal parameters
of superdiffeomorphisms and super-Lorentz transformations carrying
ghost number +2.

Analogously to~\equ{WARD_GRAV}, the Ward operator
related to~\equ{SUPERGRAV} is given by
\eq
\WW^{S}_{(\x,\th)}=\int d^{3}\!x~\sum_{\vf}
\lp\d^{S}_{(\x,\th)}\vf\rp
\frac{\d}{\d\vf} \ .
\eqn{WARD_SUPERGRAV}
After some calculations the corresponding Ward identity takes
the form
\eq
\WW^{S}_{(\x,\th)}\S=\D^{cl}_{(\x)}+\int d^{3}\!x~\x^{\l}
s(e^a_\l e_{\m a} \Xi^\m) \ ,
\eqn{WARD_SUPERGRAV_ID}
where the classical breaking writes\footnote{Let us remark, that
the breaking does not depend on the super-Lorentz parameter
$\th$ altough it is present in \equ{SUPERGRAV}.}
\eq
\D^{cl}_{(\x)}=\int d^{3}\!x~\lc -\g^{\m A}\LL_{\x}A^A_\m
+\t^A\LL_{\x}c^A -\ve_{\m\n\r}\x^\m\g^{\n A}
s(eE_a^{\r}E^{\s a}\6_{\s}\bar{c}^{A})\rc \ ,
\eqn{BREAK_CLASSIC}
and the field polynomial $\Xi^\m$ stands for
\eq
\Xi^\m=\frac{1}{2}\ve^{\m\n\r}(\6_{\n}\bar{c}^{A})
(\6_{\r}\bar{c}^{A}) \ .
\eqn{XI}

This means that the local gravitational supersymmetry is broken by two
kinds of terms. One linear in the quantum fields and a second which
is quadratic and corresponds to the hard breaking.
In the renormalization procedure, the latter needs more
attention. It can be absorbed in the original action~\equ{ACTION_1}
by coupling it to two further auxiliary fields,
namely $M_\m$ and $L_\m$, forming a BRS doublet~\cite{symanzik}
\eq
sM_\m=L_\m \ ,~~~sL_\m=0 \ .
\eqn{BRS_LM}
The corresponding action term writes
\eq
\S_{L,M}=-\int d^{3}\!x~\lp L_\m\Xi^\m-M_\m s\Xi^\m \rp \ .
\eqn{ACTION_LM}
This leads to the following modified total action
\eq
\S=\S_{CS}+\S_{ext}+\S_{gf}+\S_{L,M} \ ,
\eqn{ACTION_TOTAL}
which fulfills the Ward identity
\eq
\WW^{S}_{(\x,\th)}\S=\D^{cl}_{(\x)} \ .
\eqn{WARD_IDENTITY_FULL}
The hard breaking of~\equ{WARD_SUPERGRAV_ID} is now controlled by
the auxilliary fields $M_\m$ and $L_\m$.

The local gravitational supersymmetry transformations are given by
\eqa
\d^{S}_{(\x,\th)} L_\m\=\LL_\x M_\m + \hat{e}^a_\m e_{\n a} \x^\n
+ e^a_\m \hat{e}_{\n a} \x^\n \ , \non
\d^{S}_{(\x,\th)} M_\m\=-e^a_\m e_{\n a} \x^\n \ .
\eqan{SUPERGRAV_LM}
The contributions of these new fields $L_\m$ and $M_\m$
have to be incorporated in the Ward operators of gravity and
gravitational supersymmetry in such a way that the summation
over the fields $\vf$
in \equ{WARD_GRAV} and \equ{WARD_SUPERGRAV} includes $L$ and $M$.

Moreover, the Slavnov identity \equ{SLAVNOV_1} reads now:
\eq
\SS(\S)=\int d^{3}\!x~\lp \frac{\d\S}{\d{\g}^{\m A}}
\frac{\d\S}{\d A^A_\m}+\frac{\d\S}{\d{\t}^A}\frac{\d\S}{\d c^A}
+B^A\frac{\d\S}{\d\bar{c}^A}+\hat{e}^a_\m\frac{\d\S}{\d e^a_\m}
+L_\m\frac{\d\S}{\d M_\m}\rp = 0 \ .
\eqn{SLAVNOV_TOTAL}

We display now the resultant linear algebra obeyed by the
three operators introduced before, namely the linearized Slavnov
operator $\SS_{\S}$, the Ward operators of gravity $\WW^{gr}$ and
of gravitational supersymmetry $\WW^{S}$:
\eqa
\SS_{\S}\SS_{\S}\=0 \ , \non[4mm]
\lac \SS_{\S},\WW^{gr}_{(\e,\O)} \rac \=0 \ , \non[4mm]
\lac \WW^{gr}_{(\e,\O)},\WW^{gr}_{(\e',\O')} \rac \=
-\WW^{gr}_{\lp\lac(\e,\O),(\e',\O')\rac\rp} \ , \non[4mm]
\lac \SS_{\S},\WW^{S}_{(\x,\th)} \rac \=
\WW^{gr}_{(\x,\th)} \ , \non[4mm]
\lac \WW^{gr}_{(\e,\O)},\WW^{S}_{(\x,\th)} \rac \=
-\WW^{S}_{\lp\lc(\e,\O),(\x,\th)\rc\rp} \ , \non[4mm]
\lac \WW^{S}_{(\x,\th)},\WW^{S}_{(\x',\th')} \rac \= 0 \ ,
\eqan{ALGEBRA}

\noindent
where the new Ward operators on the r.h.s. are explicitely given
by\footnote{Remark that $\LL_{\e}\e'^\n=\{\e,\e'\}^\n$ and
$\LL_{\e}\x^\n=[\e,\x ]^\n$.}
\vspace{2mm}
\eqa
\WW^{gr}_{\lp\lac(\e,\O),(\e',\O')\rac\rp} \=
\int d^{3}\!x~\sum_{\vf}
\lac \d^{gr}_{(\e,\O)},\d^{gr}_{(\e',\O')}\rac\vf
\frac{\d}{\d\vf} \ ,\\[4mm]
\WW^{S}_{\lp\lc(\e,\O),(\x,\th)\rc\rp} \= \int d^{3}\!x~
\lc
\ve_{\m\n\r}(\LL_{\e}\x^\n)\chi^{\r A}\frac{\d}{\d A^A_\m}
-(\LL_{\e}\x^\m)A^A_\m\frac{\d}{\d c^A}
+[\LL_{\e},\LL_{\x}]\bar{c}^A\frac{\d}{\d B^A} \right. \non
&&\left. -(\LL_{\e}\x^\m)\t^A\frac{\d}{\d \g^{\m A}}
+[\d^{gr}_{(\e,\O)},\d^{gr}_{(\x,\th)}]e^a_\m
\frac{\d}{\d \hat{e}^a_\m}
- e^a_\m e_{\n a}(\LL_{\e}\x^\n)\frac{\d}{\d M_\m} \right. \non
&&\left.
+\lp[\LL_{\e},\LL_{\x}]M_\m
+ (\hat{e}^a_\m e_{\n a}+e^a_\m \hat{e}_{\n a})
(\LL_{\e}\x^\n)\rp\frac{\d}{\d L_\m}
\rc \ .
\eqan{DEF_WARD_OPER}

Starting with the gravitational invariances
of the total action, we have constructed a local gravitational
supersymmetry, which together with the Slavnov operator form a
closed linear graded algebra.


\section{Stability and Finitness}

We are now in position to address the problem  of quantizing
the theory. It is well-known, that this procedure can be achieved by
solving the following independant problems.
The stability problem of the theory leads to the discussion of the
most general deformation of the classical action induced by quantum
corrections and the anomaly problem with respect to a generalized
nilpotent symmetry operator, which is related to the validity of
the classical symmetries at the quantum level.

Let us hence start with the classical action
$$
\S = \S_{CS} + \S_{ext} + \S_{gf} + \S_{L,M} \ ,
$$
we have constructed in the previous sections as a solution of:
\label{pageconstr}
\begin{enumerate}
\item the gauge condition\footnote{Through the presence of $L$ and
$M$ the gauge condition is modified into:
\eq
\frac{\d\S}{\d B^A}=\6_\m(eE^\m_a E^{\n a} A^A_\n)
+\ve^{\m\n\r}(\6_\m M_\n)\6_\r \bar{c}^A \ .
\eqn{MOD_GAUGE_CONDITION}}~\equ{GAUGE_CONDITION} ,
\item the ghost equation\footnote{Also for the ghost equation
one has now:
\eq
\lp \frac{\d}{\d\bar{c}^A}+\6_\m(eE^\m_a E^{\n a}
\frac{\d}{\d\g^{\n A}}) \rp \S
= -\6_\m \lp s(eE^\m_a E^{\n a}) A^A_\n \rp
+\ve^{\m\n\r}\6_\m (M_\n\6_\r B^A-L_\n\6_\r \bar{c}^A) \ .
\eqn{MOD_GHOST_EQU}}~\equ{GHOST_EQU} ,
\item the Slavnov identity~\equ{SLAVNOV_TOTAL} ,
\item the Ward identitiy for gravity~\equ{WARD_GRAV_ID} ,
\item the Ward identity for gravitational
supersymmetry~\equ{WARD_IDENTITY_FULL} ,
\item the antighost equation~\equ{ANTIGHOST} .
\end{enumerate}
In order to study the influence of quantum corrections,
we now look at the most general action $\S'$, which fulfills
the same set of constraints. More precisely, we consider
$$
\S ' = \S +\D  \ ,
$$
where the perturbation $\D$ is an integrated local field polynomial
of dimension zero and ghost number zero.
The latter is constrained by:
\eqa
\fud{\D}{B^A}\= 0 \label{del-gauge} \ , \\
\lp \fud{}{\bar{c}^A}+\pa_\m ( eE^\m_a E^{\n a}
      \fud{}{\g^{\n A}})\rp\D \= 0\label{del-ghost} \ , \\
\SS_\S\D\=0 \label{del-slav} \ , \\
\WW^{gr}_{{(\e,\O)}} \D\= 0\label{del-grav} \ , \\
\WW^{S}_{{(\x,\th)}} \D\= 0\label{del-susy} \ , \\
\int d^{3}\!x~ \lp \frac{\d\D}{\d c^A} +\l f^{ABC}\cb^B
\frac{\d\D}{\d B^C}\rp  \=0 \label{del-antig} \ .
\eea

At this point we can notice that, due to the first two equations
\equ{del-gauge} and \equ{del-ghost},
the perturbation $\D$ is independent of $B$, and that it
depends on the antighost $\bar{c}$
and the external field $\g^\m$  through the combination
$\c^\m$~\equ{COMBINATION}, \ie
$\D=\D (A,c,\c,\t,e^a,\hat{e}^a,M,L)$.
Therefore, the last equation~\equ{del-antig} reduces to
$$
\int d^{3}\!x~\fud{\D}{c^A} = 0 \ .
$$

The remaining three equations can be condensed into a
single cohomology problem
\eq
\d \D =0 \ ,
\eqn{coho}
where\footnote{Let us remind that the two Ward operators
$\WW^{gr}_{{(\e,\O)}}$ and
$\WW^{S}_{{(\x,\th)}}$ carry ghost number one, just as the linearized
Slavnov operator $\SS_\S$.}
\eq
\d=\SS_\S + \WW^{gr}_{{(\e,\O)}} + \WW^{S}_{{(\x,\th)}} \ .
\eqn{full-coho}

Indeed, $\d$ can be transformed into a coboundary operator
($\d^2 =0$) if we let it act on the infinitesimal parameters
$(\e ,\x ,\th ,\O )$ in the following way:
\eqa
\d \e^\m \= \frac{1}{2}\lac \e , \e \rac^\m -\x^\m \ , \\
\d \x^\m \= \lc \e , \x \rc^\m \ , \\
\d \O^a{}_b \=\O^a{}_c \O^c{}_b - \th^a{}_b + \LL_\e \O^a{}_b \ , \\
\d \th^a{}_b \= - \th^a{}_c \O^c{}_b + \O^a{}_c \th^c{}_b
                 + \LL_\e \th^a{}_b - \LL_\x \O^a{}_b \ .
\eea
The full list of ghost numbers and form degrees is given
in Table~\ref{Dim}.
\begin{table}[hbt]
\centering
\begin{tabular}
{|l|   r|  r|  r|  r|   r|  r|  r|  r|  r|  r|  r|  r|  r|} \hline
    & $A$ & $c$  &$\tc $  &$\tilde{\t}$  &$M$  &$L$  &$e$  &$\hat{e}$
                 & $d$ &$\e$ &$\x$ &$\O$ &$\th$ \\ \hline
Dimension &1 &0 &2 &3 &-1  &-1  &0  &0  &1 & -1 &-1 & 0 & 0 \\ \hline
$\fp$ &0 &1  &-1 &-2 &1 &2  &0  &1  &0  &1 &2 &1 &2 \\ \hline
Form degree &1 &0 &2 &3 &1 &1 &1 &1 &1  &0 &0 &0 &0   \\ \hline
\end{tabular}
\caption[t1]{Dimensions, ghost numbers and form degrees.}
\label{Dim}
\end{table}

In order to solve the cohomology problem, we will use the following
general result: the cohomology of a general
non-homogenous\footnote{We speak here about homogeneity in the field,
a degree 1 is attributed to all the fields of the theory including
the parameters.} coboundary operator $\d$ is isomorphic to a
subspace of the cohomology
of its homogenous part~\cite{olivier2,dixon,piguet,becchi1}.
Therefore, we first split the operator $\d$ as
$$
\d = \d_0 +\d_1 \ ,
$$
where $\d_0$ is the part of $\d$ which does not increase the
homogeneity degree whereas $\d_1$ contains the remaining
part. One can easily check that both $\d_0$ and $\d_1$ are nilpotent
$$
\d_0^2=\lac \d_0 ,\d_1 \rac = \d_1^2 = 0 \ .
$$
In a first step, we look for the solution of the following
simplified cohomology problem:
\eq
\d_0 \D (A,c,\c,\t,e^a,\hat{e}^a,M,L) =0 \ ,
\eqn{simp-coho}
where the $\d_0$ transformations are just the homogeneity preserving
part of the transformation $\d$, that is
\eq\ba{lclclcr}
\d_0 A_\m^A \=-\pa_\m c^A \ , &\qquad &&  \\[2mm]
\d_0 c^A \= 0 \ , &&&   \\[2mm]
\d_0 \c^{\m A} \=-\e^{\m\n\r}\pa_\n A_\r^A \ , &&&  \\[2mm]
\d_0  \t^A\= -\pa_\m \c^{\m A} \ , &&&   \\[2mm]
\d_0  e_\m^a\= \hat{e}_\m^a \ ,  & \d_0 \hat{e}_\m^a \= 0 \ , \\[2mm]
\d_0  M_\m\= L_\m \ , & \d_0  L_\m\= 0 \ ,  \\[2mm]
\d_0 \e^\m  \= -\x^\m \ , & \d_0 \x^\m  \= 0    \ , \\[2mm]
\d_0 \O^a{}_b \= -\th^a{}_b \ , & \d_0 \th^a{}_b \= 0 \ ,
\ea\eqn{d-0}
where all the fields which appear in doublets are out of the
cohomology~\cite{olivier2}.
\noindent
The transformation laws for the relevant fields, in the space of
forms, are
\eq\ba{ll}
\d_0 A^A = d c^A  \ , & \d_0 c^A = 0 \ , \\[2mm]
\d_0 \tc^A =  -d A^A  \ ,~~~~~~~~
& \d_0  \tilde{\t}^A = d\tc^A \ .
\ea\eqn{d-0_FORMS}

Thus, we have to look at the integrated polynomial
$\D (A,c,\tc,\tilde{\t})$
with dimension 0 and ghost number 0:
\eq
\D =\int_\MM f^0_3 \ ,
\eqn{sol-coho}
where we use the notation $f^q_p$ for a local polynomial of form
degree $p$ and ghost number $q$.
The cohomology problem~\equ{simp-coho} is equivalent to
\eq
\d_0 f^0_3 = -d f^1_2 \ .
\eqn{hassan-0}
At this point, we use the facts that the operator $\d_0$ is
nilpotent, that it anticommutes with the external derivative $d$
and that the cohomology of $d$ is trivial
in the space of local field polynomials~\cite{olivier2,brandt}.
This allows us to write the following tower of descent equations
\eqa
\d_0 f^1_2 \= -d f^2_1 \ , \non
\d_0 f^2_1 \= -d f^3_0 \ , \label{hassan}\\
\d_0 f^3_0 \= 0 \ . \nonumber
\eea
The general solution for $f^3_0$ in~\equ{hassan}
is a multiple of $f^{ABC}c^Ac^Bc^C$
which is the only polynomial in the field which has form degree 0
and ghost number 3. Then, at the level of $f^0_3$, this gives the
following local solution
$$
f^0_3 =  \a f^{ABC}\lp -A^A A^B A^C + 6 A^A \tc^B c^C
                       + 3\tilde{\t}^A c^B c^C \rp \ ,
$$
where trivial terms of the form $(\d_0 l^{-1}_3 + d l^0_2)$ have
been droped out and $\a$ is some constant.

Therefore, at the integrated level, the solution of the homogenous
cohomology problem takes the form
\eq
\D = \a \D_c +\d_0 \tilde{\D} \ ,
\eqn{full-sol}
with
\eq
\D_c = \int_\MM f^{ABC}\lp -A^A A^B A^C + 6 A^A \tc^B c^C
                       + 3\tilde{\t}^A c^B c^C \rp  \ .
\eqn{delta-c}
At this point, we have to extend our result to the full cohomology
operator~\equ{full-coho}. It is straightforward to check that $\D_c$
also satisfies $\d\D_c=0$. Therefore, the solution of the
full $\d$ cohomology problem takes the form
\eq
\D = \a \D_c +\bar{\D} \ ,
\eqn{full-coho-sol}
with $\bar{\D}=\d \hat{\D}$.

At this step, it is more simpler to discuss the problem at the level
of $\bar{\D}$ than the one of $\hat{\D}$.
Indeed, we know that $\bar{\D}$ directly satisfies the
constraints~\equ{del-gauge}-\equ{del-antig}. The same restrictions
for $\hat{\D}$ have to be check case by case, because they strongly
depend on the coboundary operator $\d$.
Therefore, let us first look at all the possible terms with ghost
number 0 and form degree 3 one can built with the set of fields in
Table 1.
These terms are\footnote{Actually, there are further terms depending
on the parameters of the transformations which may be possible.
Their irrelevance is analyzed in the appendix.}:
\eqa
&&f^{ABC}A^A A^B A^C \ , \non[2mm]
&&f^{ABC}A^A c^B \tc^C \ , \non[2mm]
&&f^{ABC}c^A c^B \tilde{\t}^C \ , \\[2mm]
&&A^A d A^A \ , \non[2mm]
&&c^A d \tc^A \ . \nonumber
\eea
The first three terms are the non-trivial ones, their invariant
combination being $\D_c$. Therefore, the most general solution
for $\bar{\D}$ is given by
\eq
\bar{\D}=\int_\MM \lc\b_1\lp A^A dA^A\rp
+\b_2\lp c^A d\tc^A \rp\rc \ .
\eqn{contre-gen}
The parameters $\b_1$ and $\b_2$ have to be fixed to zero by imposing
the invariance of~\equ{full-coho-sol}:
\eq
\d\lp \a \D_c+ \bar{\D}\rp =
\d\int_\MM \lc\b_1\lp A^A d A^A \rp +\b_2\lp c^A d \tc^A \rp\rc=0 \ .
\eqn{final}
Then, the most general deformation of the classical action is just
$\a\D_c$ which would correspond to a continuous renormalization of
the parameter $\l$ of the model. But such a term is forbidden by the
antighost equation~\equ{del-antig}. Therefore, the classical action
admits no deformation.

The remaining part to examine concerns the anomalies.
It is well-known, that anomalies would correspond to a term of ghost
number one. Thus, in our case,
it would be of the form $\D=\int_\MM f^1_3$.
In order to determine such a term, let us use the same method as
before. That is, to look first at the non-trivial solution of the
following descent equations:
\eqa
\d_0 f^1_3 \= -d f^2_2 \ , \\
\d_0 f^2_2 \= -d f^3_1 \ , \\
\d_0 f^3_1 \= -d f^4_0 \ , \\
\d_0 f^4_0 \= 0 \ .
\eea
The only possible zero form for $f^4_0$ is
$$
f^4_0= t^{[ABCD]}c^Ac^Bc^Cc^D \ ,
$$
which is zero due to the absence of such a four rank, totally
antisymmetric, invariant tensor. Therefore, the triviality of the
$\d_0$ cohomology implies that of the full $\d$ one
and this concludes the proof of the absence of anomalies.

The consequences of the previous analysis are the following.
The absence of counterterms is responsible for the stability of the
extended classical action,
\ie the action which satisfies the constraints listed at the
beginning of this section.
This guarantees that the number of parameters of the theory remains
fixed. In such a case, the renormalizability of the theory is proved
by showing that the set of constraints are also valid at the quantum
level. It is well-known, that the gauge
condition~\equ{GAUGE_CONDITION}, the ghost equation~\equ{GHOST_EQU}
and the antighost equation~\equ{ANTIGHOST} fulfill this
requirement~\cite{blasi,rouet}. Nevertheless, in our case, we see
that the hard breaking of~\equ{WARD_SUPERGRAV_ID} enforces us to
extend our model by including the field doublet $(L,M)$.
Thus, the gauge condition and the ghost equation
are also extended to~\equ{MOD_GAUGE_CONDITION}
and~\equ{MOD_GHOST_EQU} in order to be consistent.
These modifications are linear in the quantum fields. Therefore the
resultant conditions remain renormalizable.

The remaining constraints, \ie the Slavnov and Ward identities,
due to the absence of anomalies, also survive at the quantum level.
Thus, the theory is UV finite.


\section*{Conclusions}

The Chern-Simons theory in curved space-time~\cite{olivier} was
reanalyzed in the vielbein formalism.
We have seen the explicit independence of the model with respect
to the affine spin connection and the torsion.
The main extension lies in
the additional local Lorentz symmetry and its corresponding
local supersymmetry.
The resultant graded algebra generated by the Ward operators
has the same structure, but now with the local gravity
transformations and the local gravitational supersymmetry instead of
diffeomorphisms and superdiffeomorphisms respectively.

Although the additional exact symmetries require two
further parameters, they do not change the stability property.
They are also free of anomalies.

Then the UV finiteness follows, provided the existence of a
consistent substraction scheme. This require that the manifold
must be topologically equivalent to a flat space and possess
an asymptotically flat metric.

\vspace{1cm}
{\bf Acknowledgements:}

We are grateful to M.W. de Oliveira and O. Piguet
for useful discussions and suggestions.


\section*{Appendix: Analysis of the trivial part of the counterterms}

\setcounter{equation}{0}
\renewcommand{\theequation}{A.\arabic{equation}}

The most general deformation of the classical action is given by
\eq
\D=\a\D_c+\bar{\D}~~~,~~~\bar{\D}=\d\hat{\D} \ ,
\eqn{mo-1}
where $\D$ is an integrated local polynomial in the fields
with dimension 0 and ghost number 0,
which is the solution of the full cohomology problem for the operator
$\d$ defined in \equ{full-coho}:
\eq
\d\D=0 \ .
\eqn{COHO-PROB}
The non-trivial part $\D_c$ has already
been determined \equ{delta-c}.
The perturbation $\D$ and therefore also $\bar{\D}$ cannot
depend on the fields $(\e,\O)$ and $(\x,\th)$,
since they are nothing else than infinitesimal
parameters of the corresponding field transformations.
Furthermore, both have to fulfill the constraints \equ{del-gauge}
and \equ{del-ghost}. This means that
$\bar{\D}=\bar{\D}(A,c,\c,\t,e^a,\hat{e}^a,M,L)$.
But, due to the fact that the full cohomology operator $\d$
\equ{full-coho} is nilpotent only if it acts on the parameters,
it could be possible that $\hat{\D}$ depends on the parameter fields.

In the following, we want to discuss the trivial part of the solution
of the cohomology problem at the level of $\hat{\D}$. To do this, we
are looking first to the validity of the constraints \equ{del-gauge}
and \equ{del-ghost} for $\hat{\D}$.

Let us introduce the bosonic filtering operator
\eq
\NN=\int d^{3}\!x~\lp B^A\frac{\d}{\d B^A}
+ \bar{c}^A\frac{\d}{\d \bar{c}^A} \rp \ .
\eqn{FILTER}
Due to the independence of $\D$ with respect to $B$ \equ{del-gauge}
and $\bar{c}$ \equ{del-ghost}, this implies that
\eq
\NN\D=0 \ .
\eqn{FILTERING}
If we define now the fermionic operator
\eq
\RR=\int d^{3}\!x~\lp \bar{c}^A\frac{\d}{\d B^A} \rp \ ,
\eqn{OPERATOR}
it follows from \equ{del-gauge} that $\RR\D=0$.
Therefore, the anticommutator between $\d$ and $\RR$ acting on $\D$
yields
\eq
\lac \d , \RR \rac \D = \NN\D \ .
\eqn{ANTICOMM}
Here, it is important to consider the counterterm $\D$ and not the
full action $\S$. Indeed, if we act with the same combination on
$\S$, we would get some contributions from the
breaking in \equ{WARD_IDENTITY_FULL},
but for the counterterm $\D$ such breaking is not present. Therefore,
using the Jacobi identity, we get the vanishing commutator
\eq
\lc \NN, \d \rc \D = 0 \ ,
\eqn{COMM}
which is precisely the starting point for the general procedure.
Following~\cite{brandt}, it is clear that the expansion
of $\D$ in terms of eigenvectors of $\NN$
\eq
\D=\sum_{n=0}^{\bar{n}} \D^{(n)}~~~,~~~\NN\D^{(n)}=n\D^{(n)} \ ,
\eqn{EWG}
leads to
\eq
\d\D^{(n)}=0 \ .
\eqn{DELTA-N-1}
With the equations \equ{mo-1}, \equ{FILTERING} and \equ{COMM}
one finds
\eq
\NN\bar{\D}=0=\NN\d\hat{\D}=\d\NN\hat{\D}
=\d\sum_{n=1}^{\bar{n}} n\hat{\D}^{(n)} \ ,
\eqn{DELTA-N-2}
which implies
\eq
\d\hat{\D}^{(n)}=0 ~~~~~ \forall n \ge 1 \ .
\eqn{EXP}
Therefore, the only contribution to the counterterm has the form
\eq
\bar{\D}=\d\hat{\D}^{(0)} \ ,
\eqn{COUNTER}
with $\NN\hat{\D}^{(0)}=0$. This means that also $\hat{\D}$ is
independent of the fields $B$ and $\bar{c}$, but they can depend on
the parameter fields,
i.e. $\hat{\D}=\hat{\D}(A,c,\c,\t,e^a,\hat{e}^a,M,L,\e,\O,\x,\th)$.

Indeed, for the complete set of possible terms with form
degree 3 and ghost number 1 to $\hat{\D}$ one has:
\eq
\ba{ll}
{}~~~~~&  \L_1=A^A \tilde{\chi}^A \ , \\[2mm]
                   &  \L_2=c^A \tilde{\t}^A \ , \\[2mm]
{}~~~~~&  \L_3=i_\e A^A \tilde{\t}^A \ , \\[2mm]
             &  \L_4=i_\e \tilde{\chi}^A \tilde{\chi}^A \ , \\[2mm]
{}~~~~~&  \L_5=i_\e i_\e \tilde{\chi}^A \tilde{\t}^A \ , \\[2mm]
{}~~~~~&  \L_6=i_\e i_\e i_\e \tilde{\t}^A \tilde{\t}^A \ ,
\ea
\eqn{LIST}

\noindent
where $i_\e$ denotes the inner derivative with respect to $\e$
(e.g. $i_\e A^A = \e^\m A^A_\m$).

At the integrated level
\eq
\hat{\D}_i =\int_{\MM}\L_i \ ,~~~~~i=1,...,6
\eqn{FORM_Z}
the action of the operator $\d$ yields
\eqa
\d\hat{\D}_1\=\int_{\MM}\lc - A^A dA^A + (dc^A)\tilde{\chi}^A
+ (i_\x A^A)\tilde{\t}^A
+ (i_\x \tilde{\chi}^A)\tilde{\chi}^A
\rc \ , \label{mo1} \\
\d\hat{\D}_2\=\int_{\MM}\lc - (i_\x A^A)\tilde{\t}^A
-(dc^A)\tilde{\chi}^A
\rc \ , \label{mo2} \\
\d\hat{\D}_3\=\int_{\MM}\lc - (i_\x A^A)\tilde{\t}^A
+(i_\e dc^A)\tilde{\t}^A + (i_\e i_\x \tilde{\chi}^A)\tilde{\t}^A
- (i_\e A^A)d\tilde{\chi}^A
\rc \ , \label{mo3} \\
\d\hat{\D}_4\=\int_{\MM}\lc -(i_\x \tilde{\chi}^A)\tilde{\chi}^A
+2(i_\e i_\x \tilde{\chi}^A)\tilde{\t}^A
-2(i_\e dA^A)\tilde{\chi}^A
\rc \ , \label{mo4} \\
\d\hat{\D}_5\=\int_{\MM}\lc
-2(i_\e i_\x \tilde{\chi}^A)\tilde{\t}^A
-(i_\e i_\e i_\x \tilde{\t}^A)\tilde{\t}^A
-(i_\e i_\e \tilde{\chi}^A)d\tilde{\chi}^A
-(i_\e i_\e dA^A)\tilde{\t}^A
\rc \ , \label{mo5} \\
\d\hat{\D}_6\=\int_{\MM}\lc
-3(i_\e i_\e i_\x\tilde{\t}^A)\tilde{\t}^A
-2(i_\e i_\e i_\e \tilde{\t}^A)d\tilde{\chi}^A
\rc \ . \label{mo6}
\eqan{margot}
The last expression $\d\hat{\D}_6$ cannot contribute to $\bar{\D}$,
because it has no corresponding term, which could cancel the
parameter dependence of the second term on the r.h.s. of \equ{mo6}.
Now this implies that also $\d\hat{\D}_5$ is forbidden due to the
second term in \equ{mo5}. Analogously, owing to the second term in
\equ{mo3}, we have to reject $\d\hat{\D}_3$, which demands the
non-consideration of $\d\hat{\D}_4$ caused by the second term.
Finally, from the fourth term on the r.h.s. of \equ{mo1}
follows that one has to drop $\d\hat{\D}_1$
and therefore also the last expression $\d\hat{\D}_2$ does not
contribute to $\bar{\D}$.
Thus, we have shown the irrelevance of the terms
\equ{mo1}-\equ{mo6}.

At the end, let us remark that in our case the discussion of
possible counterterms at the level of $\bar{\D}=\d\hat{\D}$, as done
in section 4, is much simpler than at the one of $\hat{\D}$.


\end{document}